\documentclass[%
 reprint,
 amsmath,amssymb,
 aps,
]{revtex4-2}

\usepackage{graphicx}
\usepackage{dcolumn}
\usepackage{bm}
\usepackage{float}
\usepackage{xcolor}
\usepackage{soul} 
\usepackage{siunitx}

\begin{document}

\title{Mott vs Kondo: \\ Influence of Various Density Functional Based Methods on the Ce Isostructural Phase Transition Mechanism}

\author{Brenden W. Hamilton}
\email{brenden@lanl.gov}
\author{Alexander R. Muñoz}
\author{Travis E. Jones}
\author{Benjamin T. Nebgen}
\affiliation{Theoretical Division, Los Alamos National Laboratory, Los Alamos, New Mexico 87545, USA}

\date{\today}

\begin{abstract}

The cerium iso-structural phase transition ($\gamma$ to $\alpha$) is dominated by f-electron localization changes that results in a magnetic ordering change and a volume collapse.
Generally, these physics are difficult to capture with ab initio and first principles methods.
However, previous works have shown various methods to be successful in predicting at least some of the physics of the $\gamma$ to $\alpha$ phase transition.
Therefore, here, we perform a broad survey of density functional based methods across three levels of theory and types of functions (GGA, MetaGGA, and Hybrid functionals)
and compare the results, focusing on hydrostatic compression across the phase boundary at zero Kelvin.
For the methods that best reproduce experimental results, we directly probe the predicted mechanisms and frame the results in the Mott/Kondo debate,
assessing how the underlying methods and assumptions of different functionals can assess the physical drivers in the phase transition,
providing insight into the governing dynamics of this unique phase transition.

\end{abstract}

\maketitle

\section{Introduction}

Materials under pressure can undergo a wide range of unique and exotic phase and state 
transformations\cite{Reed2008TransientSemimetallic,chao1962stishovite,Kadau2002Microscopic,katayama2000first,hamilton2024high,sastry2003liquid,hamilton2025thermal}.
When possible, the use of first principles and theory to describe these events is the optimal modeling effort\cite{zhang2024exploring,kapil2022first,gati2021pressure,sanchez1999ab,serrano2004pressure,f2025improving}.
Face centered cubic (FCC) cerium is known to undergo a pressure induced iso-structural phase transformation from the $\gamma$ to $\alpha$ phase, which is characterized experimentally by a volume collapse 
\cite{ramirez1971theory,beecroft1960existence,decremps2011diffusionless,lipp2008thermal,pickett1981local}.
The first principles description of this transition, and \textit{f}-electron materials in general, is considered a significant challenge and the optimal method(s) are highly debated.

For cerium, the $\gamma$ phase is ferromagnetic while the $\alpha$ phase exhibits Pauli paramagnetism\cite{rocher1962etude,macpherson1971magnetic}, 
and there is a discrete change in properties across the volume collapse, despite having the same atomic structure\cite{lipp2017anomalous,moore2011watching,decremps2011diffusionless,perdew2009workhorse}.
The governing mechanism of this phase transition is highly debated between the Mott transition\cite{johansson1974alpha,johansson1995calculated} and Kondo volume collapse\cite{allen1982kondo,allen1992alpha,zhu2020kondo}.
A variety of other theories, such as Zachariasen-Pauling theory, Coqblin-Blandin model, Ramirez-Falicov model, and the Hirst model, have also been put forward\cite{koskenmaki1978cerium,coqblin1968stabilite,ramirez1971theory,hirst1974configuration}.
The Mott picture assumes a change from localized to delocalized \textit{f} electrons due to a competition between the intersite \textit{f-f} hybridization and on-site \textit{f-f} Coulomb interactions\cite{casadei2016density}.
Conversely, the Kondo case states that \textit{spd} hybridization with the \textit{f} electrons causes a change in screening with decreased volume.
Continued work has shown that these two mechanisms are not necessarily separable and both models may have roles to play in capturing the underlying physics of this unique and complex mechanism\cite{amadon2006alpha,lanata2013gamma,bieder2014thermodynamics}.
Lastly, the volume collapse has been shown to extend into the liquid phase\cite{cadien2013first} and previous works have indicated that thermal disorder and entropy play a significant role in the transition at finite temperatures\cite{amadon2006alpha,jarlborg2014role}.

In the literature, \textit{ab initio} techniques have been used previously to probe the interactions necessary to drive structural and magnetic phenomena\cite{kunes_2008,kasinathan_2007, gaifutdinov_2024, lee_2025}.
The attempts to describe the interplay of the underlying physical mechanisms of these phenomena have lead to a variety of attempts to study Ce with \textit{ab initio} techniques.
In studies using density functional theory, specifically the local density approximation (LDA) and the generalized gradient approximation (GGA), 
there is a strong delocalization of the f-electrons, leading to a stabilized $\alpha$ phase, independent of volume\cite{wang2008thermodynamics}.
However, GGA+U with spin polarization has been able to capture the $\gamma$ phase as well\cite{tran2014nonmagnetic}, where non-magnetic calculation only capture the $\gamma$ phase with no double minimum in the cold curve.
Hybrid functionals, such as HSE06 or PBE0, have been shown to capture both phases and the double minimum of the cold curve in a single methodology\cite{casadei2012density}.
These hybrid DFT calculations predict that the driving mechanisms for these transitions occur at zero temperature\cite{casadei2016density,lanata2013gamma}.
Using Hybrid DFT with a random-phase approximation, the correct phase ordering is predicted, where the $\alpha$ phase is lower in energy, however, the predicted minimum energy volumes are far from expected values\cite{casadei2012density}.
Additionally, dynamical mean field theory and quantum Monte Carlo techniques have been used to study the electronic origin of the volume collapse in Ce\cite{devaux_2015,lanata2013gamma}.
However, in hybrid  DFT, the phase ordering is incorrect and the computational cost of using other methods is high for more complex tasks like training machine learned interatomic potentials (MLIPs).

Here, we utilize several levels of theory to study the $\gamma$ to $\alpha$ phase transition in the interest of identifying low-cost techniques for MLIPs.
We use DFT at several levels of approximation, including GGAs, MetaGGAs, and hybrid functionals to identify which methods correctly predict the isostructural transition ordering and the magnetic phase transition as a function of pressure.
By comparing the predictions of each method, we can frame the necessary interactions in DFT that are needed to capture the isostructural transition from $\gamma$-Ce to $\alpha$-Ce.
We then frame these interactions in terms of the Mott and Kondo mechanisms underlying the volume collapse of Ce.
We find that the Hybrid functionals and PBE+$U_{ff}$ (Hubbard U acting on the f orbitals) with a $U_{ff}$ value greater than 2 eV lead to Mott like mechanisms, whereas MetaGGA functionals and intermediate values of $U_{ff}$ (1 to 2 eV) result in
mechanisms more aligned with the Kondo volume collapse.
The results of the Kondo volume collapse mechanism methods align well with experimental understanding and high-level theoretical techniques,
especially that the $\alpha$ phase is energetically preferable at zero Kelvin and the equilibrium volumes in both phases.

\section{Methods}
The calculations presented in this work were performed in the projector augmented wave method as implemented in the Vienna Ab initio Simulation Package (VASP)\cite{Kresse1996PlaneWave}.
To examine the interactions at each level of theory in DFT, we employ GGA, meta-GGA, and hybrid functional approximations.
In the GGA, we use PBE and PBE+$U_{ff}$ functionals where the Hubbard U value is varied to control the degree of 4\textit{f} localization in the system\cite{Perdew1996GGA,anisimov1991diffusion}. 
At the Meta-GGA level, we use R2SCAN, R2SCAN-L, and rTPSS functionals \cite{furness2020accurate,kaplan2022laplacian,mejia2017deorbitalization}.
This provides different functionals that incorporate the kinetic energy density ($\tau$) and the Laplacian of the density. 
The rTPSS functional is included to have one non-SCAN based functional.
At the hybrid functional level, we use HSE06, PBE0, B3LYP, and SCAN0\cite{krukau2006influence,ernzerhof1999assessment,stephens1994ab,hui2016scan,gerber2005hybrid}.
This provides a range of functional types, having screened and unscreened interactions with HSE06 and PBE0 (respectively), a non-PBE based hybrid functional with B3LYP, and
a MetaGGA based hybrid functional with SCAN0.
For all studies, we use the crystallographic 4-atom fcc cell with a 400 eV kinetic energy cutoff, a $6\times6\times6$ $\Gamma$-centered {\bf k}-point mesh, and an energetic convergence criteria of 1E-8 eV.

We assume collinear spin polarization and initialize the magnetism separately on each site 1.1 $\mu_B$, 1.2 $\mu_B$, 1.3 $\mu_B$, and 1.4 $\mu_B$.
We refer to this as a symmetry-broken ferromagnetic (FM) state.
The symmetry-broken FM state aids the energetic convergence and prevents stabilization into local magnetic minima in the energy landscape.
Each method is used to calculate the electronic state of the crystal for lattice parameters ranging from 4.6 \AA to 5.3 \AA, where the experimental $\gamma$ and $\alpha$ lattice parameters are 5.16 \AA 
and 4.81 \AA \cite{koskenmaki1978cerium,lipp2008thermal}, with the volume collapse happening near the $\alpha$ equilibrium volume.

In several approximations, there are nearly degenerate states with different magnetic moments near the phase transition.
To identify the low energy state at the relevant volumes, we ran sweeps of the magnetic moments with a spacing of 0.1 $\mu_B$ on each atom from 0.0 $\mu_B$ to 1.2 $\mu_B$ to identify the magnetic structure with the minimum energy. 
Additional points are added as needed to find the energy minimum point with a spacing of 0.01 $\mu_B$.
As this methodology is cost-prohibitive for the hybrid functionals, we employ a four-point search: the symmetry-broken FM state, a constrained to 0.0 $\mu_B$ state, a constrained 1.0 $\mu_B$ per site state, and a calculation with unconstrained magnetic moments of initial magnitudes of 2.1 $\mu_B$, 2.2 $\mu_B$, 2.3 $\mu_B$, and 2.4 $\mu_B$. 

\section{Results and Discussion}
\subsection{GGA Results: PBE and PBE+U}

While PBE and other GGA methods have historically been workhorse functionals for metallic systems, they are generally not as accurate when \textit{f}-electron physics comes into play\cite{anisimov1991diffusion}.
The inclusion of a Hubbard U correction enables better accuracy for highly correlated electronic systems such as localized \textit{f}-electrons\cite{liechtenstein1995density,dudarev1998electron}.
However,  U must be parametrized, often using a physical quantity such as the equilibrium volume or more sophisticated models such as the linear response function\cite{moore2024high}.
In the first section, we assess the accuracy of PBE and PBE+$U_{ff}$ for the prediction of the Ce cold curve and for the FCC-FCC iso-structural phase transition ($\gamma$ to $\alpha$), as well as the magnetic evolution, where the larger
volume $\gamma$ phase is FM and the compressed $\alpha$ phase exhibits Pauli-paramagnetism.

Figure 1 shows the energy in reference to each curves minimum as a function of lattice parameter for PBE and PBE+$U_{ff}$ with several values of $U_{ff}$.
For the PBE calculations, in blue, the curve does not exhibit a double minimum that is the primary signature of the isostructural phase transition.
In Figure 2, we show the total magnetization for each of these calculations for the various PBE and PBE+$U_{ff}$ for four values of U.
While PBE exhibits FM behavior at large lattice parameters, as is experimentally expected, PBE fails to capture the localized \textit{f}-electrons which prevents the isostructural phase transition.
PBE does, however, accurately predict the equilibrium volume for the $\alpha$ phase. 

\begin{figure}[htpb]
  \includegraphics[width=0.4\textwidth]{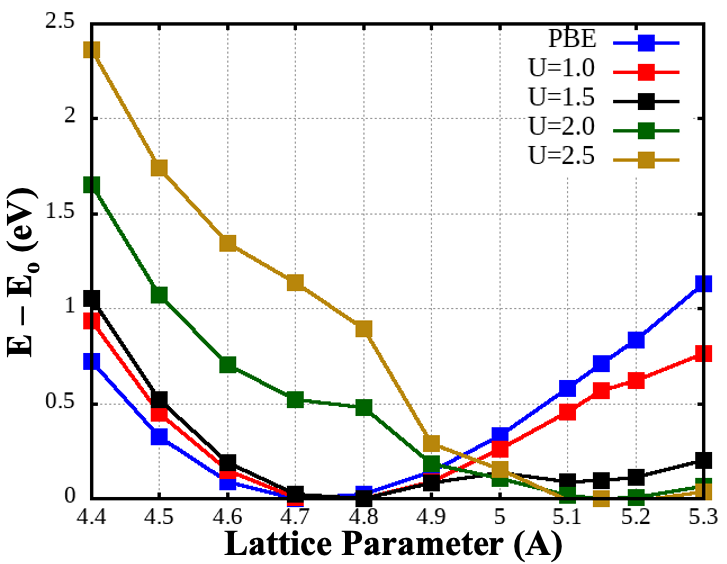}
  \caption{The energy in reference to the minimum as a function of lattice parameter for PBE and PBE+U with four values for U.}
  \label{fig:Fg1}
\end{figure}

With a Hubbard U correction applied in Figure 1 and 2, we begin to see the formation of a double-minimum in the cold curve.
For U = 1.0 eV, only the $\alpha$ phase is observed, but with deviations from PBE at large volumes.
As the Hubbard U is increased to larger values, the expected double minimum occurs.
Additionally, in Figure 2, for $U_{ff}$ = 1 eV with the delocalized \textit{f}-electrons, the total magnetization evolves slowly, transitioning continuously between the FM and paramagnetic states, whereas there is a more distinct drop in magnetization for higher $U_{ff}$ values.

For $U_{ff}$ greater than 1 eV, we see an energetic evolution where the $\alpha$ phase is lower in energy for $U_{ff}$ = 1.5 eV, whereas the $\gamma$ phase is lower in energy for $U_{ff}$ = 2.0 eV and 2.5 eV.
While the $\gamma$ phase is energetically favorable at room temperature, the $\alpha$ phase is lower in energy at zero Kelvin. 
Experimentally, the transition from $\alpha$ being lower in energy to $\gamma$ being the preferred phase at ambient pressure occurs at ~116 K\cite{gschneidner1962effects}.
Only $U_{ff}$ = 1.5 eV predicts both the double minimum and the correct energetic ordering of the phases. 
As $U_{ff}$ increase over 2.0 eV, we see the energetic gap increase and become discontinuous.
From Figure 2, we see that $U_{ff}$ = 1.5 eV predicts a discrete magnetic phase transition, whereas larger values of U have points near a total magnetization (4 atoms) of 2 $\mu B$.
These points are antiferromagnetic state with three up moments and one down. 
This is likely due to the high value of U overstabilizing the magnetic ordering.

In all three cases where $U_{ff}$ $\geq$ 1.5 eV, the lattice parameter at the two minima are very close to the experimental equilibrium values.
Overall, we find that a PBE+$U_{ff}$ functional with $U_{ff}$ = 1.5 eV leads to the best total prediction of phase ordering, magnetic moments, and equilibrium volumes.

\begin{figure}[htpb]
  \includegraphics[width=0.4\textwidth]{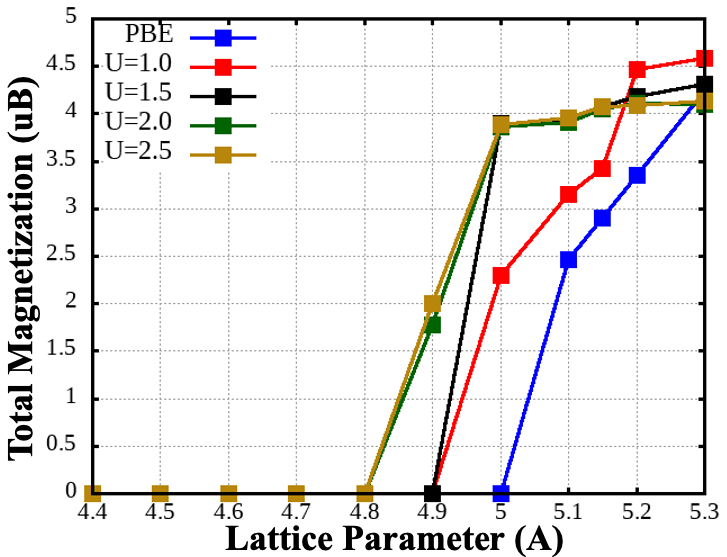}
  \caption{Total magnetization as a function of lattice parameter for PBE and PBE+$U_{ff}$ with several values for $U_{ff}$.}
  \label{fig:Fg2}
\end{figure}

\subsection{MetaGGA Results}

Turning now to MetaGGA functionals, we repeat the same analysis as the GGA functionals for R2SCAN, R2SCAN-L, and revTPSS.
Two of the three (R2SCAN and R2SCAN-L) are based on the SCAN functional\cite{sun2015strongly}, which obeys all 17 known exact constraints for Meta-GGA functionals
with good computational scaling.
R2SCAN\cite{furness2020accurate} is an improved version of SCAN that maintains the exact constraints and transferability, while increasing computational efficiency.
R2SCAN-L\cite{kaplan2022laplacian,mejia2017deorbitalization} is a Laplacian-modified version of R2SCAN that improves convergence rates and prevents over-magnetization\cite{mejia2020meta}.
The Revised Tao-Perdew-Staroverov-Scuseria (revTPSS)\cite{perdew2009workhorse} functional expands on TPSS\cite{tao2003climbing} by restoring the density gradient expansion for exchange over a wide range of densities,
leading to more accurate material properties such as lattice constants, surface energies, and atomization energies.

Figure 3 shows the energy in reference to the minimum as a function of lattice parameter for all three MetaGGA functionals.
Here we see that revTPSS does not predict the $\gamma$ phase, but gives a reasonable equilibrium volume for the $\alpha$ phase.
Both of the SCAN based functionals do predict the phase transition, and both correctly predict the energetic ordering, with the $\alpha$ phase at lower energy.
R2SCAN-L has a smaller energy difference between these phases.
For equilibrium lattice parameter, R2SCAN is off by over 0.1 \AA for both phases (predicting 4.7 \AA and 5.0 \AA compared to experimental 4.81 \AA and 5.16 \AA),
whereas R2SCAN-L accurately predicts the lattice parameters.

\begin{figure}[htpb]
  \includegraphics[width=0.4\textwidth]{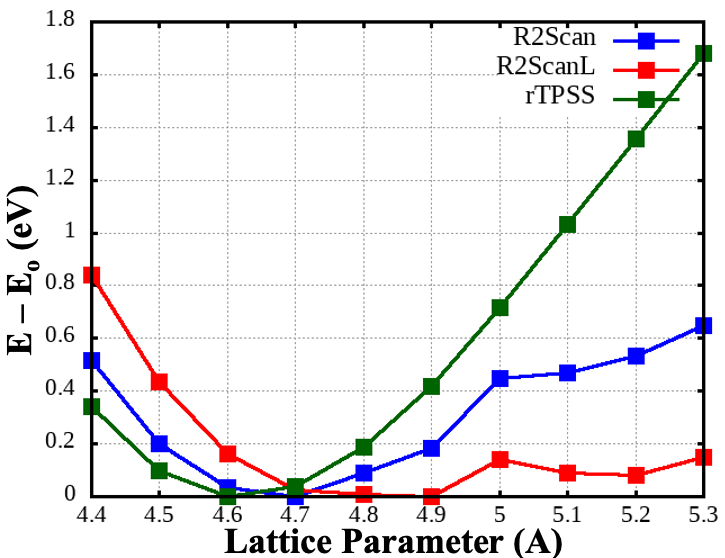}
  \caption{The energy in reference to the minimum as a function of lattice parameter for the MetaGGA functionals.}
  \label{fig:Fg3}
\end{figure}

Figure S-1 in the Supplemental Material shows the total magnetization for each of these calculations for all three MetaGGA functionals.
All predict a continued rising magnetization with compression of the $\gamma$ phase, especially past the equilibrium volume.
With no functional distinguishing itself as more accurate based on the magnetic moment predictions, we find that the R2SCAN-L functional makes the overall best predictions across the 
$\gamma to \alpha$ phase transition in pure Ce.
The difference in the R2SCAN and R2SCAN-L results can be explained as the improvements in the functional from adding the 
Laplacian term in a deorbitalized version which is typically better at predictions of localized states and defects where regular SCAN can underestimate charge localizations\cite{mejia2017deorbitalization}.


\subsection{Hybrid Functional Results}

We now apply the same analysis to a variety of hybrid functionals.
We employ HSE06, PBE0, B3LYP, and SCAN0.
HSE06 is a range separated functionals, whereas PBE0, SCAN0, and B3LYP are unscreened hybrid functionals.

The PBE0 functional\cite{ernzerhof1999assessment} employs a 0.25 Hartree-Fock exchange and 0.75 PBE exchange functional: $E_{xc}^{PBE0} = \frac{1}{4} E_x^{HF} + \frac{3}{4}E_x^{PBE} + E_c^{PBE}$.
SCAN0\cite{hui2016scan} also follows this formalism with:  $E_{xc}^{SCAN0} = \frac{1}{4} E_x^{HF} + \frac{3}{4}E_x^{SCAN} + E_c^{SCAN}$, using a MetaGGA functional instead of PBE.
The HSE06 functional\cite{krukau2006influence} follows the same form, mixing PBE and HF exchange, but only in the short range (SR) part and PBE only in the long range (LR) part where the screening length is 0.2 $A^{-1}$:
$E_{xc}^{HSE06} = \frac{1}{4} E_x^{HF,SR} + \frac{3}{4}E_x^{PBE,SR} + E_x^{PBE,LR} + E_c^{PBE}$.
Lastly, the unscreened B3LYP functional\cite{stephens1994ab} combines the exchange (consisting of 80\% of the LDA exchange term plus 20\% of the Hartree-Fock exchange term, with 72\% of the gradient corrections of the Becke88 exchange functional\cite{becke1988density}) 
and correlation terms (consisting of 81\% of LYP correlation energy and 19\% of the Vosko-Wilk-Nusair correlation functional\cite{vosko1980accurate}):
$E_x^{B3LYP} = 0.8E_x^{LDA} + 0.2E_x^{HF} + 0.72E_x^{B88}$ and $E_c^{B3LYP} = 0.19E_c^{VWN3} + 0.81E_c^{LYP}$.

Figure 4 shows the energy in reference to the minimum as a function of lattice parameter for the four hybrid functionals.
Each case results in the prediction of the iso-structural phase transition, but with the incorrect energetic ordering and significantly larger energy increases than in the GGA and MetaGGA functionals.
PBE0 predicts the volume collapse at a lattice parameter smaller than expected from experiment ($\approx$ 4.6 \AA), whereas HSE06 and B3LYP are much closer to the expected value of 4.8 \AA.
Interestingly, SCAN0 predicts an antiferromagnetic state from L = 4.6 \AA to 4.9 \AA, leading to the three distinct curves in Figure 4. These states are energetically more favorable than a FM or zero moment structure.

\begin{figure}[htpb]
  \includegraphics[width=0.4\textwidth]{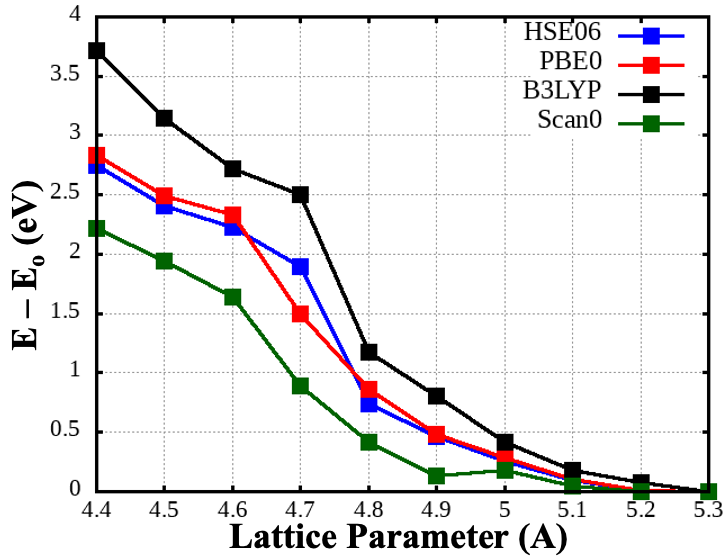}
  \caption{The energy in reference to the minimum as a function of lattice parameter for all Hybrid functionals used here.}
  \label{fig:Fg5}
\end{figure}

Figure S-2 in the Supplementary Material shows the total magnetization for each of these calculations for the tested hybrid functionals.
These results track with the cold-curves, where PBE0 over-stabilizes the FM state at small volumes and most cases stabilize an antiferromagnetic state at a minimum of one volume,
where there are 3 positive and one negative moment in the HSE06 and B3LYP cases, and zero total magnetization in the SCAN0 case.
Interestingly, these results, in the frame of volumes, phase ordering, and magnetic transformations, are not overly accurate compared to the ``lower'' levels of theory, though the hybrid functionals are associated increased computational cost.
In the following section, we will directly compare and more deeply assess the best cases from each level of theory: PBE+$U_{ff}$ ($U_{ff}$=1.5 eV), R2SCAN-L, and HSE06.


\subsection{Comparison Across Levels of Theory}

To compare the different levels of theory, we begin by assessing the cold curves.
Figure 5 shows the energy in reference to the minimum as a function of lattice parameter for PBE+$U_{ff}$ ($U_{ff}$ = 1.5 eV and 2.5 eV), R2SCAN-L, and HSE06.
There are two regimes in the predictions.
HSE06 and PBE+$U_{ff}$ for $U_{ff}$ = 2.5 eV predict distinct results from $U_{ff}$ = 1.5 eV and R2SCAN-L, where each of the two sets give similar results.
$U_{ff}$ = 1.5 eV and R2SCAN-L have the minimum energy in the $\alpha$ phase and a very small energy difference between the local minimums.
$U_{ff}$ = 2.5 eV and HSE06 have the minimum energy in the $\gamma$ phase and have a large energy difference between the phases, with the $\alpha$ curve minimum being directly at the phase transition volume.
Previous work has shown that adding a random phase approximation (RPA) to PBE0 calculations can be used to get the correct phase ordering with hybrid functionals, but results in much poorer volume predictions\cite{casadei2012density}.
Due to the increased computational cost and adding another independent variable, we do not explore the affect of RPA here.

\begin{figure}[htpb]
  \includegraphics[width=0.4\textwidth]{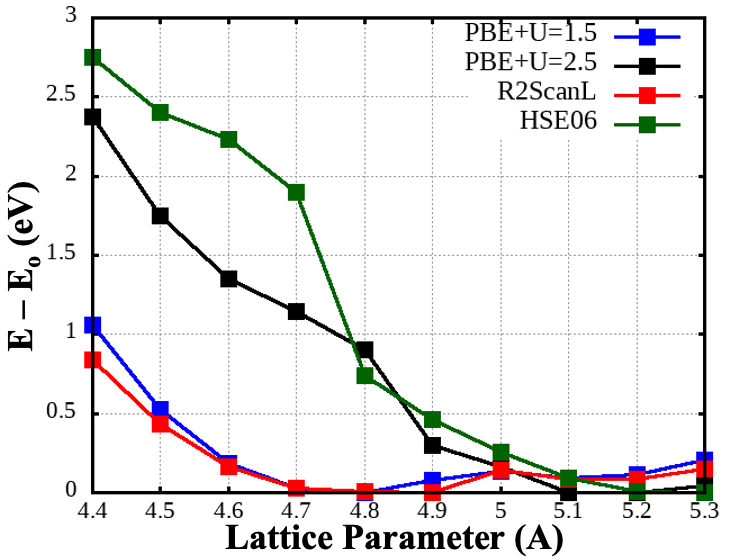}
  \caption{The energy in reference to the minimum as a function of lattice parameter for most accurate functionals from each level of theory.}
  \label{fig:Fg7}
\end{figure}

From here, we look into the predicted properties and mechanisms across the phase transition.
Figure 6 shows the predicted density of states for each method for the f orbitals at a range of different lattice parameters.
From this we see, again, a distinct difference between two classes of results where HSE06 and PBE+$U_{ff}$ for $U_{ff}$ = 2.5 eV predicts very different results from $U_{ff}$ = 1.5 eV and R2SCAN-L.
Supplementary Material Figure S-3 shows the density of states (referenced to each system's Fermi energy) for L = 4.80 and 5.15 \AA ($\alpha$ and $\gamma$) from each level of theory for the f and d orbitals.

For HSE06, starting with the L = 5.2 \AA FM state, we see highly localized {\it f}-orbitals, with the two peaks over 5 eV apart.
All of the hybrid functionals, including HSE06, in Section V, overstabilize the $\gamma$ phase, leading to the incorrect energetic ordering, where the localized, FM state is
at a much lower energy that the non-magnetic $\alpha$ phase. Hence, these functionals may be overlocalizing the {\it f} electrons, leading to the large peak in Figure 6.
Comparatively, the same incorrect energetic ordering and strong localization occurs in PBE+$U_{ff}$ for large values ($U\geq2$ eV) of the Hubbard U, which also drives localization in the {\it f} states.
However, the $U_{ff}$ = 2.5 eV case shown here does not separate the localized unoccupied and occupied peaks in the FM case as much as HSE06.

\begin{figure*}[htpb]
  \includegraphics[width=0.7\textwidth]{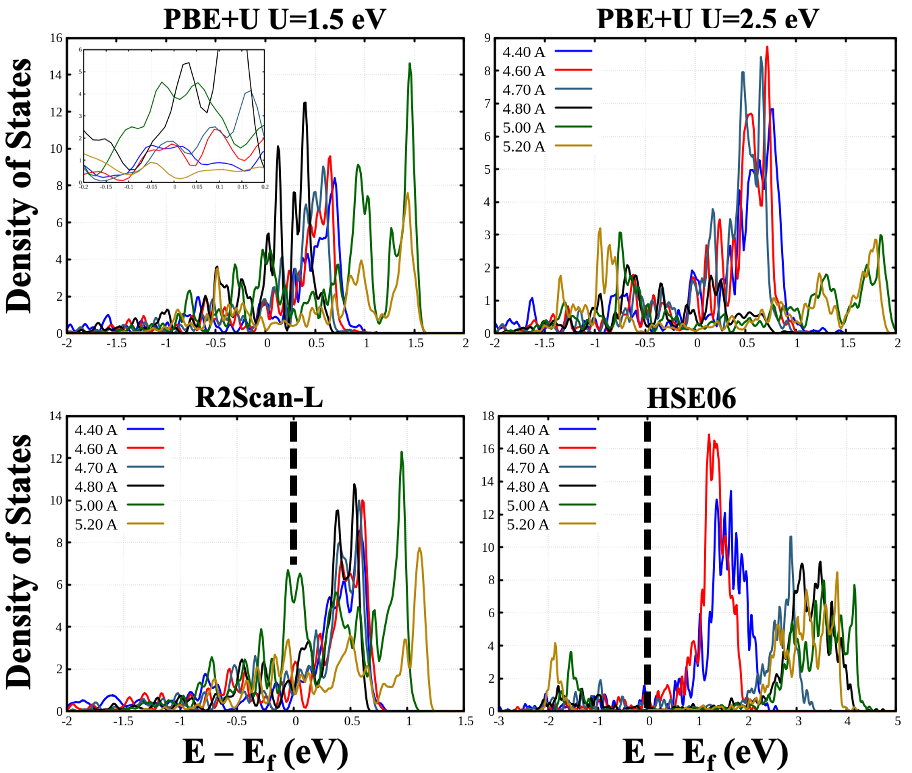}
  \caption{The density of states referenced to each system's Fermi energy for a range of lattice parameters from each level of theory.}
  \label{fig:Fg8}
\end{figure*}

In the non-magnetic case (e.g. L = 4.6 \AA), we see relatively few f-states below the Fermi energy.
As we compress from the FM states at larger volumes towards the phase transitions, we see the lower energy {\it f} states move toward the Fermi energy and lowering in magnitude,
eventually crossing over at the phase transition point where almost all {\it f} states are above the Fermi energy in a single peak.
After the phase transition, the few {\it f} states below the Fermi energy are strongly hybridized by the {\it d} states, (see Supplementary Figure S-3).
This leads to the hybrid functionals predicting a Mott-like mechanism for the phase transition in FCC Ce, where bringing the atoms closer together causes orbital overlap and delocalization of the available {\it f} states.

Conversely, the PBE+$U_{ff}$ ($U_{ff}$=1.5 eV) and R2SCAN-L functionals paint a much different picture. The bands shown in Figure 6 are narrower and show higher levels of available {\it f} states
in the $\alpha$ phase.
While peaks tend to move toward the Fermi energy with compression, there is never a distinct jump as shown in the HSE06 DOS,
and a continuum of states still exist across the Fermi energy (also very strongly hybridized with the {\it d} orbitals as per SM Figure S-3).
The U=1.5 eV panel of Figure 6 shows a zoom in of the same plot at the Fermi energy, showing DOS peaks directly at the Fermi energy
\cite{kondo1964resistance}. 
The same features exist in the R2SCAN-L DOS.

For zero to low values of U correction on PBE, we see an absence of the $\gamma$ phase altogether.
While a FM state does appear at larger lattice parameters, no double minimum in the cold curve appears.
Whereas HSE06 and R2SCAN-L fundamentally predict different mechanisms, by tuning the Hubbard U parameter in PBE+$U_{ff}$, we can control the predicted mechanism, albeit at the cost of accuracy in other areas.
The Mott mechanism predicted cold curves get the energetic ordering incorrect for zero temperature, and predict a potentially unphysical energy jump between the two phases.
This points toward a preference toward the Kondo mechanism, as expected from higher levels of theory like quantum Monte Carlo.

\section{Conclusions}

Overall, we find that a range of methods of the density functional flavor can be used to model the $\gamma$ to $\alpha$ phase transition in pure Ce.
While methods at different levels of theory can produce reasonable predictions of the FCC Ce cold curve, the underlying mechanisms driving the predictions vary.
The hybrid functionals such as HSE06 predict a strong localization of the {\it f}-electron where almost all states are pushed above the Fermi energy at the phase transition.
This likens to a Mott transition mechanism driving the phase transition.
We see the same fundamental mechanism for PBE+$U_{ff}$ with $U_{ff} \geq 2$ eV, where the strong Hubbard U term penalizes the non-localized $\alpha$ phase,
promoting energetic stability of the ferromagnetic $\gamma$ phase.

For PBE+$U_{ff}$ with lower values for $U_{ff}$ and MetaGGA functionals like R2SCAN-L, we see a fundamentally different picture.
The result in a mechanism more aligned with the Kondo volume collapse picture.
As the Hubbard U is lowered more in PBE+U, this leads to no phase transition at all and only predictions of the $\alpha$ phase.

It is the results of the Kondo volume collapse mechanism methods that most aligns with experimental understanding,
especially that the $\alpha$ phase is energetically preferable at zero Kelvin and for predicting the equilibrium volumes in both phases.
This does not inherently mean that the Kondo mechanism acts alone, but that spectrum of mechanistic response controlled by the Hubbard U in PBE+$U_{ff}$
may point to the fact that the Kondo and Mott pictures are not distinct and that the Cerium volume collapse may exist in some overlapping state between them.

Overall, there are a wide range of DFT functionals that can be appropriate for studying cerium, depending on the necessities of the specific study, especially where computational cost in concerned. 
This present study paves the way for the development of a Ce (and potentially other lanthanide materials) training set for 
MLIPs based upon various quantum mechanical methodologies for modeling dynamic properties.

\section{Acknowledgments}

Funding for this project was provided by the Advanced Simulation and Computing Physics and Engineering Models project (ASC-PEM). Partial funding was provided by
ASC Computational Systems \& Software Environment (CSSE).
This research used resources provided by the Los Alamos National Laboratory Institutional Computing Program, which is supported by the U.S. Department of Energy National Nuclear Security Administration under Contract No. 89233218CNA000001.
Approved for Unlimited Release LA-UR-25-30010.

\bibliography{references_1017}

\end{document}